\documentclass
[twocolumn,aps,pra,superscriptaddress,groupaddress]{revtex4-1}
\usepackage[latin9]{inputenc}
\usepackage{amsmath}
\usepackage{amssymb}
\usepackage{graphicx}

\makeatletter
\@ifundefined{textcolor}{}
{%
 \definecolor{BLACK}{gray}{0}
 \definecolor{WHITE}{gray}{1}
 \definecolor{RED}{rgb}{1,0,0}
 \definecolor{GREEN}{rgb}{0,1,0}
 \definecolor{BLUE}{rgb}{0,0,1}
 \definecolor{CYAN}{cmyk}{1,0,0,0}
 \definecolor{MAGENTA}{cmyk}{0,1,0,0}
 \definecolor{YELLOW}{cmyk}{0,0,1,0}
}

\usepackage{bm}\usepackage{amsthm}\usepackage{epstopdf}\usepackage{txfonts}

\DeclareRobustCommand{\openzero}{\leavevmode\hbox{0\kern-.55em0}}

\makeatother

\begin{document}
\title{Measures of coherence generating power for quantum unital operations}

\author{Paolo Zanardi, Georgios Styliaris, Lorenzo Campos Venuti}

\affiliation{Department of Physics and Astronomy, and Center for Quantum Information
Science \& Technology, University of Southern California, Los Angeles,
CA 90089-0484}

%
%
%
%
%
%

\begin{abstract}
Given an orthonormal basis in a $d$-dimensional Hilbert space and a unital quantum operation $\cal E$ acting on it
one can define  a non-linear mapping that associates to $\cal E$  a  $d\times d$ real-valued matrix that we call the Coherence Matrix
of $\cal E$ with respect to $B$.  We show that one can use this coherence  matrix to define vast families of measures of the coherence generating power (CGP) of the operation.
These measures have a natural geometrical interpretation as separation of $\cal E$ from the set of incoherent unital operations.
The probabilistic approach to CGP  discussed in  P. Zanardi et al., arXiv:1610.00217 can be reformulated and generalized
introducing, alongside the coherence matrix, another $d\times d$ real-valued matrix, the Simplex Correlation Matrix. This matrix describes the relevant statistical correlations
in the input ensemble of incoherent states.  Contracting these two matrices one finds CGP measures associated with the process of preparing the given incoherent ensemble and processing it
with the chosen unital operation. Finally, in the unitary case, we discuss how these concepts  can be made compatible with an underlying tensor product structure
by defining families of CGP measures that are additive.
\end{abstract}
\maketitle
\section{Introduction}
Well defined phase relationships  between different branches of the wavefunction and the associated interference effects is one of the distinctive
features of quantum systems.  This phenomenon, closely related to the fundamental superposition principle, is known as {\em{quantum coherence}}
and it is nowadays known to be one the basic elements necessary for quantum information processing \cite{Nielsen-Book}. More recently quantum coherence
has been recognized as key ingredient for quantum metrology \cite{Iman-ArXiv-2016}, quantum biology  \cite{Q-Biology} and quantum thermodynamics \cite{Lostaglio-prx-2015,Lostaglio-nat-comm-2015}.

This state of affairs  spurred a renewed  great interest  in the {\em{quantitative}} theory of coherence \cite{Baumgratz-prl-2014,Adesso-review}.
The latter, in turn, can be  regarded as special case of a broader set of {\em{resource}} theories of {\em{asymmetry}}  \cite{Iman-pra-2014,Iman-nat-comm-2014,Iman-Thesis-2012}.
Given a specific ``resource'' e.g., entanglement,  one defines the set of quantum states that are resource-free e.g., separable states, and the set of ``cheap'' operations
e.g., LOCC transformations, namely quantum maps that are unable to enhance the amount of resource contained in the states they are operating on.

Once these distinguished sets are defined, one can move on to establish a quantitative theory of the resource by introducing suitable real-valued functions that measure,
in some sense, the degree of separation of states from the resource-free ones.  Similarly, the quantification of the ability of a quantum operation i.e.,
completely positive (CP) map, to generate resource can be seen as a measure of its separation from the cheap operations as well as an indication of its usefulness
in suitable quantum protocols e.g., teleportation.

In the case of quantum coherence different approaches have been discussed in the literature as to quantify the coherence
of a quantum state \cite{Iman-ArXiv-2016,Baumgratz-prl-2014,Levi-NJP-2014, Girolami-prl-2014,Yao-sci-rep-2016} as well as the
coherence generating power (CGP)  of quantum operations \cite{CP1,CP2,CP3}. For a thoughtful and comprehensive review see Ref. \cite{Adesso-review}.

In \cite{CGP_0},  following the spirit of Ref.~\cite{zanardi-praRC-2000} in entanglement theory, we proposed
a probabilistic  approach to CGP. Given any preferred basis $B$  in the Hilbert space of the system, we defined the CGP of a map as the average coherence that is produced
by the  quantum operation acting on a suitable   input ensemble of incoherent states. This approach leads to fully  analytical  expressions for CGP of quantum
maps in any dimensions and can be related to simple operational protocols \cite{CGP_0}. However, it is in a sense specifically tailored for unital CP maps
and based on a maximally symmetric, essentially uniform,  input ensemble of incoherent states.

\begin{figure}[t]
\begin{centering}
\includegraphics[width=0.9\columnwidth]{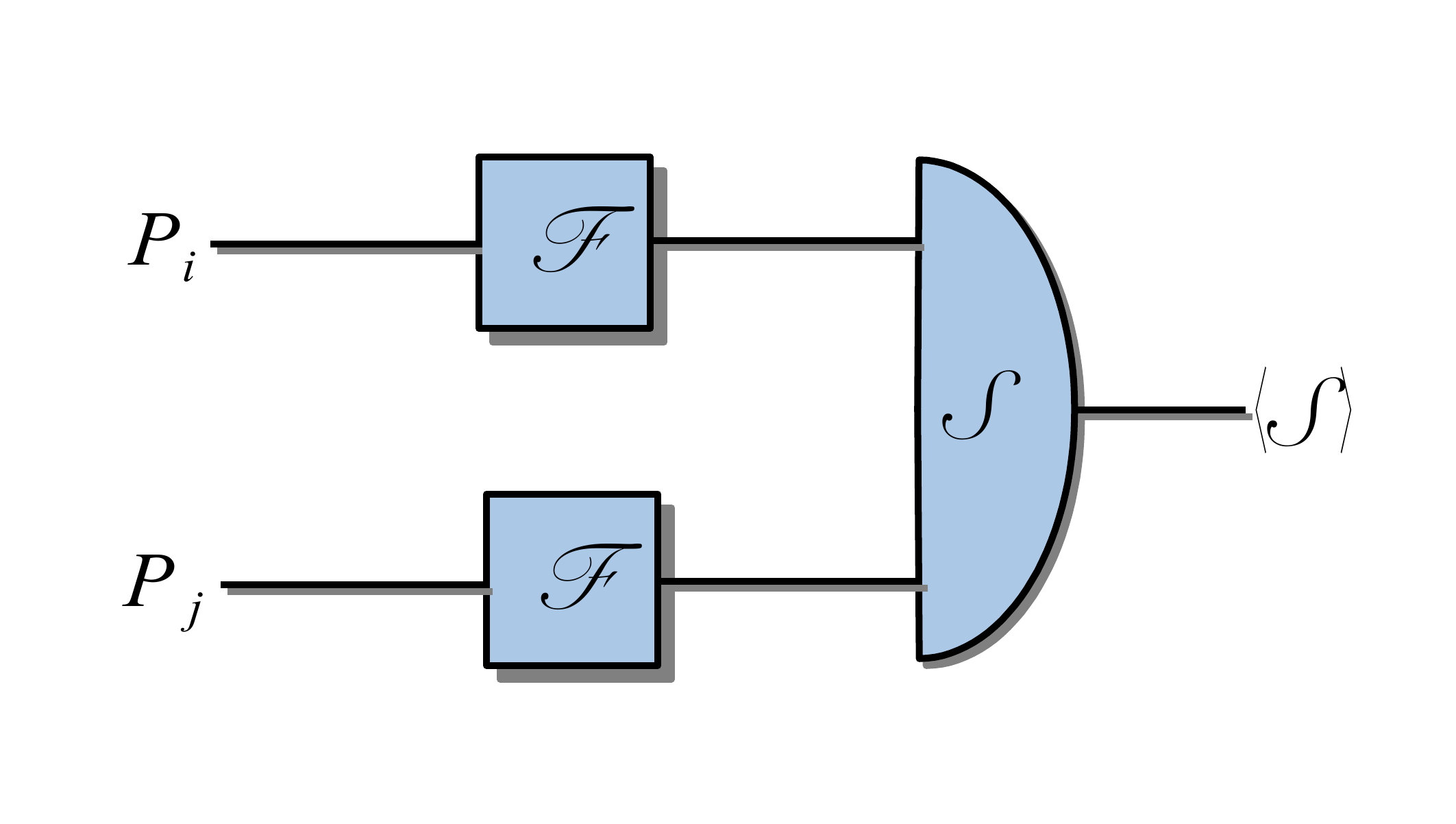}
\par\end{centering}
\vspace{-4mm}
\caption{Protocol for the direct detection of the  matrix $\hat{A}_{ij}({\cal F})$  of the linear map $\mathcal{F}$ according to Eq.~\eqref{A-protocol}. The system is initialized in the state $P_i \otimes P_j $, sent through the channel $\mathcal{F}^{\otimes 2}$,  the swap operator $S$ is measured and its expectation value is obtained. The procedure must be repeated for all the $d(d+1)/2$ pairs of indices $i\le j$.
\label{protocol-Fig}}
\end{figure}

In this paper we considerably generalize the results of \cite{CGP_0}  by introducing novel families of analytical CGP measures for {\em{unital}} maps
that are independent of any statistical input ensemble as well as allowing more general ensembles. This is achieved by ``deconstructing" the approach
in \cite{CGP_0} in terms of its basic independent ingredients: the so-called {\em{coherence matrix}} of the map and the so-called {\em{simplex correlation matrix}}.
The former depends just on $B$ and the map itself whereas the latter just on $B$ and the input ensemble.

We will discuss how these measures, in the unitary case, can be interpreted in a elegant geometric fashion as ``distances" between the map and the set of cheap incoherent operations.
Moreover we will discuss the interplay between the CGP of a map and the tensor product structures that are present in the case of bi-partite quantum systems.
In this case we will introduce families of CGP measures that are {\em{additive}}.

In Sect.~II we will introduce the basic concepts and set the notation. In Sect.~III will define the coherence matrix of a unital operation and show how it can be used to
define CGP measures. In Sect.~IV we will specialize to the important case of unitary operations and discuss a geometric approach to CGPs.
In Sect.~V we will introduce the simplex correlation matrix of an input ensemble of incoherent states. In Sect.~VI we will analyze the interplay
of CGP measures and tensor product structures. Finally in Sect.~VII conclusions and outlook are provided.

\section{Setting the stage}
In this section we provide the basic definitions and set up the notation.
Let $B:=\{|i\rangle\}_{i=1}^d$ be an orthonormal basis of the Hilbert space ${\cal H}\cong{\mathbf{C}}^d$ and $\{ P_i :=|i\rangle\langle i|\}_{i=1}^d$ the associated family of one-dimensional
orthogonal projectors.  We consider the operator space $ {\mathrm{L}}({\cal H})$ over $\cal H$ as a Hilbert space equipped by the Hilbert-Schmidt scalar product
$\langle X, Y\rangle:={\mathrm{Tr}}(X^\dagger Y)$ and norm $\|X\|_2:=\sqrt{\langle X, X\rangle}$. $\Vert X \Vert_1$ denotes the (Schatten) 1-norm of operator $X$ i.e., the sum of the singular values of $X$.

All the completely positive maps considered in this paper, unless otherwise specified, are assumed to be trace-preserving
and {\em{unital}}.
\vskip 0.2truecm
{\bf{Definition 1}} 
The $B$-dephasing map ${\cal D}_B\colon {\mathrm{L}}({\cal H})\mapsto  {\mathrm{L}}({\cal H})$ is defined by
 ${\cal D}_B(X):=\sum_{i=1}^d P_i XP_i=\sum_{i=1}^d P_i {\mathrm{Tr}}(P_i X)$.
We also define  ${\cal Q}_B:={\mathbf{1}}-{\cal D}_B$.
\vskip 0.2truecm
The map  ${\cal D}_B$  (${\cal Q}_B$) is an orthogonal  projection in the Hilbert-Schmidt space  ${\mathrm{L}}({\cal H})$ and its range is the space of $B$-diagonal operators (orthogonal complement thereof).
The set of $B$-diagonal {\em{states}} form a $(d-1)$-dimensional simplex that will be denoted by $I_B\cong\{p\in{\mathbf{R}}^d\,/\, p_i\ge 0,\, \sum_{i=1}^d p_i=1\}$. The elements of $I_B$ will be referred to as {\em{incoherent}} states.
The $d$ extremal points of $I_B$ (which correspond to the projections $P_i$) will be denoted by $e_i\,(i=1,\ldots, d)$ and as elements of ${\mathbf{R}}^d$ fulfill $(e_i)_l=\delta_{il},\,(l=1,\ldots, d)$.
The center of $I_B$ is $(1/d,\ldots,1/d)$ corresponding to the maximally mixed state $d^{-1}\openone$.
\vskip 0.2truecm
{\bf{Definition 2}} 
A CP map ${\cal E}\colon {\mathrm{L}}({\cal H})\mapsto  {\mathrm{L}}({\cal H})$ is called $B$-incoherent iff $[{\cal E},{\cal D}_B]=0$.
We shall denote the set of $B$-incoherent  maps over $\cal H$ by $CP_B({\cal H})$.
\vskip 0.2truecm
An incoherent map $\cal W$ fulfills the relation ${\cal Q}_B{\cal W}{\cal D}_B=0$ \cite{example-georgios}.
This in turn implies that $I_B$ is invariant under the action of $\cal W$.
The latter property is equivalent to incoherence, as given by Def. 2, if ${\cal W}$ is (Hilbert-Schmidt) normal \cite{CGP_0}.
Let us consider an incoherent unitary CP map ${\cal W}(\bullet)=W\bullet W^\dagger, \,(W\in U({\cal H}))$.
From \cite{CGP_0} one knows that $W=\sum_{i=1}^d \eta_i |\sigma_W(i)\rangle\langle i|$ where
$\sigma_W\in{\cal S}_d$ and $\{\eta_i\}_i\subset U(1)$.  The permutations $\sigma\in{\cal S}_d$ are unitarily represented in ${\cal H}\cong{\mathbf{C}}^d$ equipped with the basis $B$ by $\sigma\mapsto P_\sigma:=\sum_{i=1}^d  |\sigma(i)\rangle\langle i|$. The only ${\cal S}_d$-invariant vector is
\begin{equation}
|\phi^+\rangle=\frac{1}{\sqrt{d}} \sum_{i=1}^d |i\rangle \label{phi^+}.
\end{equation}
It is useful to remind here that a matrix $\hat{X}\in M({\mathbf{R}}, d)$ is row (column) {\em{stochastic}} iff $ \hat{X}|\phi^+\rangle=|\phi^+\rangle$ ($ \hat{X}^T|\phi^+\rangle=|\phi^+\rangle$).
A {\em{bi-stochastic}} matrix is one that is both row and column stochastic. An example of the latter class is the projection $|\phi^+\rangle\langle\phi^+|$.

\vskip 0.2truecm
{\bf{Definition 3}}
A function $C_B\colon {\cal E}\mapsto C_B({\cal E})\in{\mathbf{R}}_0^+$ over CP maps is a Coherence Generating Power (CGP) measure with respect to the basis $B$ iff the following
properties hold. If  ${\cal W}\in CP_B({\cal H})$ then: {\bf{i)}} $C_B({\cal W})=0$ and  {\bf{ii)}} $C_B({\cal W}{\cal E})\le C_B({\cal E})$.
Moreover, if $C_B({\cal E})=0$ implies that $\cal E$ is incoherent then the CGP measure is called {\em{faithful}}.
\vskip 0.2truecm
In \cite{CGP_0} we have defined a probabilistic CGP measure by the following expression
\begin{equation}
C_B({\cal E}) ={\mathbf{E}}_\psi\left[\|{\cal Q}_B{\cal E}{\cal D}_B(|\psi\rangle\langle\psi|)\|_2^2\right], \label{CGP-Haar}
\end{equation}
where the expectation ${\mathbf{E}}_\psi[\bullet]:=\int d\mu(\psi) [\bullet]$ is performed with respect to  the Haar measure $d\mu(\psi)$. If $\rho\in I_B$ one can write $\rho=\sum_{i=1}^d p_i P_i$ where the $p_i\ge 0$'s comprise a probability distribution
($\sum_{i=1}^d p_i=1$). One has that $\|{\cal Q}_B{\cal E}(\rho)\|_2^2=\langle \sum_{i=1}^d p_i {\cal Q}_B{\cal E}(P_i), \sum_{j=1}^d p_j {\cal Q}_B{\cal E}(P_j)\rangle=
\sum_{i,j=1}^d p_i p_j \langle{\cal Q}_B{\cal E}(P_i), {\cal Q}_B{\cal E}(P_j)\rangle$.  One can then write the measure (\ref{CGP-Haar}) as
\begin{equation}
C_B({\cal E})=\sum_{i,j=1}^d  \hat{C}_{ij}({\cal E}) \hat{S}_{ji}(\mu_{Haar})=\langle \hat{C}({\cal E}), \hat{S}(\mu_{Haar})\rangle,
\label{C-S}
\end{equation}
where $ p_i(\psi):=|\langle\psi| i\rangle|^2$, $\hat{C}({\cal E})_{ij}:=
\langle{\cal Q}_B{\cal E}(P_i), {\cal Q}_B{\cal E}(P_j)\rangle$ and
$\hat{S}_{ij}(\mu_{Haar}):= {\mathbf{E}}_\psi[ p_i(\psi)  p_j(\psi)],$ $(i,j=1,\ldots,d)$.

These relations show that in the CGP approach we pursued in \cite{CGP_0} there are two independent key ingredients:
the matrices $\hat{C}({\cal E})_{ij}$ and $\hat{S}(\mu_{Haar})$. The first one depends just on the operation ${\cal E}$ while the second depends just
on the statistical correlations in the input ensemble of incoherent states.

In this paper we will analyze different ways to introduce CGP measures expanding on this simple consideration.

\section{The coherence matrix of a unital operation}
In this section we introduce the first building block for constructing vast families of CGP measures independently of the input ensemble: the coherence
matrix of a  unital quantum operation. In order to do so it is convenient to first introduce an intermediate object and establish its properties.
\vskip 0.2truecm
{\bf{Definition 4}}
Let ${\cal F}$ be a linear map   from ${\mathrm{L}}({\cal H})$ into itself mapping hermitian operators onto hermitian operators. We consider the associated $d\times d$ matrix $\hat{A}({\cal F})\in M(d, {\mathbf{R}})$
\begin{equation}
\hat{A}_{ij}({\cal F}):=\langle{\cal F}(P_i), {\cal F}(P_j)\rangle,\quad (i,j=1,\dots,d), \label{A-Matrix}
\end{equation}
which is nothing but the Gram matrix of the map ${\cal F}$ restricted to the linear (real) subspace of  ${\mathrm{L}}({\cal H})$ spanned by the $P_i$'s. Note that $\hat{A}({\cal F})=\hat{A}({\cal F}{\cal D}_B)$.
\vskip 0.2truecm
Since  ${\mathrm{Tr}}\,(AB)={\mathrm{Tr}}\left(S (A\otimes B)\right)$
where $S\colon {\cal H}^{\otimes\,2}\mapsto  {\cal H}^{\otimes\,2}$ is the {\em{swap}} operator i.e., $S|i j\rangle=|j i\rangle,\,(i,j=1,\ldots,d)$,
one can write
\begin{equation}
\hat{A}_{ij}({\cal F})={\mathrm{Tr}}\left( {\cal F}(P_i) {\cal F}(P_j)\right)={\mathrm{Tr}}\left( S {\cal F}^{\otimes\,2} (P_i\otimes P_j)\right).
\label{A-protocol}
\end{equation}
This equation suggests an operational protocol to measure the $d(d+1)/2$ independent real-valued entries of the matrix $\hat{A}({\cal F})$ associated to a CP map $\cal F$:

{\bf{Protocol}}

{\bf{0)}} Prepare product states $P_i\otimes P_j,\, (j\ge i=1,\ldots, d).$

{\bf{1)}} Perform the quantum operation ${\cal F}\otimes{\cal F}$ on each of them.

{\bf{2)}} Measure the expectation value of the swap operator on the so obtained states.
\vskip 0.2truecm
This protocol is depicted in Fig.~\ref{protocol-Fig}. We now list the main  properties of the matrix $\hat{A}$.
\vskip 0.2truecm
{\bf{Proposition 1}}

{\bf{i)}}  $\hat{A}({\cal F})=\hat{A}({\cal F})^T\ge 0$,

{\bf{ii)}} $\hat{A}({\cal F})=0\Leftrightarrow {\cal F}{\cal D}_B=0$,

{\bf{iii)}} $\|\hat{A}({\cal F})\|\le \| {\cal F}\|_{2,2}^2$, where $\| {\cal F}\|_{2,2}:=\sup_{\|X\|_2=1}\|{\cal F}(X)\|_2$.

{\bf{iv)}} If $\cal F$ is unital then $\hat{A}({\cal F})$ is bi-stochastic and $\|\hat{A}({\cal F})\|\le 1$.

{\bf{v)}} The mapping $\hat{A}\colon {\cal F}\mapsto \hat{A}({\cal F})$ is convex.

{\bf{vi)}} If $\cal W$ is unital then $\hat{A}({\cal W}{\cal F})\le \hat{A}({\cal F}).$ For $\cal W$ unitary one has
$\hat{A}({\cal W}{\cal F})=\hat{A}({\cal F}).$
\vskip 0.2truecm
{\em{Proof.--}}
{\bf{i)}} Since the $P_j$'s are hermitian and $\cal E$ is hermiticity preserving the ${\cal E}(P_j)$ are hermitian therefore
$\hat{A}_{ij}({\cal F}):=\langle{\cal F}(P_i), {\cal F}(P_j)\rangle=\langle{\cal F}(P_j), {\cal F}(P_i)\rangle=\hat{A}_{ji}({\cal F}) \forall i,j$
from which follows the symmetry property. Moreover $\sum_{i,j=1}^d v_i v_j \langle{\cal F}(P_i), {\cal F}(P_j)\rangle=\|{\cal F}(V)\|_2^2\ge 0$
where $V=\sum_{i=1}^d v_i P_i.$
{\bf{ii)}}
${\cal F}(P_i)={\cal F}{\cal D}_B(P_i)$ therefore ${\cal F}{\cal D}_B=0$ clearly implies $\hat{A}({\cal F})=0.$ Conversely if $\hat{A}({\cal F})=0$
one has that (see above) that $\|{\cal F}(X)\|_2^2=0$ for any $X=\sum_{i=1}^d x_i P_i$ namely $\cal F$ restricted to Im ${\cal D}_B$ is identically vanishing.
{\bf{iii)}} $\|\hat{A}({\cal F})\|=\sup_{\|v\|=1}\langle v, \hat{A}({\cal F}) v\rangle=\sup_{\|v\|=1} \|{\cal F}(V)\|_2^2\le \sup_{\|v\|=1} \|{\cal F}\|_{2,2}^2 \|V\|_2^2\le \|{\cal F}\|_{2,2}^2$
as $\|V\|_2^2=\sum_{i=1}^d v_i^2=\|v\|=1.$
{\bf{iv)}} $\sum_{i=1}^d \hat{A}({\cal F})_{ij}=\langle {\cal F}(\sum_{i=1}^d P_i), {\cal F}(P_j)\rangle=\langle {\cal F}(\openone), {\cal F}(P_j)\rangle={\mathrm{Tr}}\, {\cal F}(P_j)={\mathrm{Tr}}\, P_j=1, \forall j$.
This shows that $\hat{A}({\cal F})$ is column stochastic, from symmetry i.e., {\bf{i)}} bi-stochasticity follows. Moreover for all unital $\cal F$'s
one has $\|{\cal F}\|_{2,2}\le 1;$ therefore from {\bf{iii)}} one has that $\|\hat{A}({\cal F})\|\le 1$.
{\bf{v)}}
 In order to check $\hat{A}(p {\cal E}_1+q {\cal E}_2)\le p \hat{A}({\cal E}_1) +q \hat{A}({\cal E}_2),\,(p,q\ge 0, p+q=1)$
one has to check that $\langle v, X v\rangle\ge 0\,(\forall v\in{\mathbf{R}}^d, \|v\|=1$) where $X= p \hat{A}({\cal E}_1) +q \hat{A}({\cal E}_2)-\hat{A}(p {\cal E}_1+q {\cal E}_2)$.
Since $\langle v, \hat{A}( {\cal E} )v\rangle=\|{\cal E}(V)\|_2^2\ge 0,$ the desired  inequality follows from convexity of the $2$-norm.
{\bf{vi)}} $\langle v, [\hat{A}({\cal F}) -\hat{A}({\cal W} {\cal F})] v\rangle=\|{\cal F}(V)\|_2^2-\|{\cal W}{\cal F}(V)\|_2^2\ge 0,$ where the last inequality follows from $\|{\cal W}\|_{2,2}\le 1$
which holds for all unital $\cal W$'s. When $\cal W$ is unitary  equality holds.
$\hfill\Box$\vskip 0.2truecm
{\bf{Remark 1}}
In the following of the paper property {\bf{vi)}} will play a crucial role. Since the matrix $\hat{A}({\cal E})$ is defined in terms of the Hilbert-Schmidt
scalar product, property {\bf{vi)}} does not hold for general i.e., non unital, CP maps \cite{Perez-Garcia-jmp-2006}. This is the crucial point where the unitality constraint shows up.
\vskip 0.2truecm
We now  define one of the most important concepts of this paper. It is the basic building block used to the define families of CGP measures
for unital quantum operations.
\vskip 0.2truecm
{\bf{Definition 5}} Given the unital  CP map $\cal E$ its {\em{coherence matrix}} with respect to the basis $B$ is given by
\begin{equation}
\hat{C}({\cal E}):=\hat{A}({\cal Q}_B{\cal E})= \hat{A}({\cal E})-\hat{A}({\cal D}_B{\cal E})
\label{C-Matrix}
\end{equation}
%
The second equality in Eq.~(\ref{C-Matrix}) can be checked by using ${\cal Q}_B={\mathbf{1}}-{\cal D}_B$ and $\langle X, {\cal D}_B(Y)\rangle=\langle{\cal D}_B(X), {\cal D}_B(Y)\rangle$.
\vskip 0.2truecm
{\bf{Remark 2}} Notice  that if $\cal W$ is an incoherent unitary CP map one has $\hat{C}({\cal W}{\cal E})=\hat{A}({\cal Q}_B{\cal W}{\cal E})=\hat{A}({\cal W}{\cal Q}_B{\cal E})=
\hat{A}({\cal Q}_B{\cal E})$ where we have used the definition of incoherent CP map and {\bf{vi)}} of Prop. 1. This means that the coherence matrix of a CP map
is invariant by {\em{post-processing}} with incoherent unitaries.
\vskip 0.2truecm
The next proposition is one the of key ones of this paper as it summarizes the basic properties of the coherence matrix associated with a unital quantum operation $\cal E$.
It also shows how $\hat{C}({\cal E})$  can be  used to build vast families of CGP measures.
\vskip 0.2truecm
{\bf{Proposition 2}}

{\bf{i)}}  Suppose $p\colon  M(d, {\mathbf{R}})_+\mapsto{\mathbf{R}}_0^+$ is such that  a) $p({\mathbf{0}})=0$, b) $X\ge Y\Rightarrow
p(X)\ge p(Y),$  then the function ${\cal E}\mapsto p(\hat{C}({\cal E}))$ is a good CGP measure.

{\bf{ii)}} Moreover if $p(W X W^{-1})=p(X)$ for all unitary $W$'s
then $p(\hat{C}({\cal E}{\cal W}))=p(\hat{C}({\cal E}))$ where ${\cal W}(\bullet)=W\bullet W^\dagger$ is unitary  in $CP_B$.

{\bf{iii)}} $\sum_{i,j=1}^d \hat{C}_{ij}({\cal E})= d\langle\phi^+|\hat{C}({\cal E})|\phi^+\rangle=0$. 

{\bf{iv)}} The mapping  which associates to each unital map $\cal E$ its coherence matrix $\hat{C}({\cal E})$ is convex
and $0\le \hat{C}({\cal E})\le {\openone}$

{\bf{v)}} $\|\hat{C}({\cal E})\|_1={\mathrm{Tr}}\,\hat{C}({\cal E})\le d-1$.
\vskip 0.2truecm
{\em{Proof.--}}
{\bf{i)}}
 From {\bf{ii)}} of Prop. 1 one has that $\hat{C}({\cal E})=\hat{A}({\cal Q}_B{\cal E})=0$ iff ${\cal Q}_B{\cal E}{\cal D}_B=0.$  The latter property is fulfilled by  $B$-incoherent  ${\cal E}$'s.
Moreover if $[{\cal W},\,{\cal D}_B]=0$  then using {\bf{vi)}} of Prop. 1 one has
$\hat{C}({\cal W}{\cal E})=\hat{A}( {\cal Q}_B {\cal W}{\cal E})=\hat{A}(  {\cal W}{\cal Q}_B{\cal E})\le \hat{A}({\cal Q}_B{\cal E})=\hat{C}({\cal E}).$
Monotonicity of $p$ now implies $p( \hat{C}({\cal W}{\cal E}))\le p(\hat{C}({\cal E}));$
this shows monotonicity under {\em{post-processing}} by any unital incoherent ${\cal W}$.
{\bf{ii)}}  One can check that $\hat{C}({\cal E}{\cal W})=P_W^{-1} \hat{C}({\cal E}) P_W$ where $P_W$ is the permutation associated to $\cal W$ (see {\bf{i)}} in Prop. 8). Thereby
$p(\hat{C}({\cal E}{\cal W}))=p(P_W^{-1} \hat{C}({\cal E}) P_W)=p(\hat{C}({\cal E})).$
{\bf{iii)}}
$\sum_{i,j=1}^d \hat{C}_{ij}({\cal E})=\sum_{i,j=1}^d \langle {\cal Q}_B{\cal E}(P_i), {\cal Q}_B{\cal E}(P_j)\rangle=\|{\cal Q}_B{\cal E}(\openone)\|_2^2=\|{\cal Q}_B(\openone)\|_2^2=0$.
{\bf{iv)}} Convexity can be proven in the same way as in {\bf{v)}} in Prop. 1. Moreover from {\bf{iv)}} of Prop. 1 follows that $\|\hat{C}({\cal E})\|=\|\hat{A}({\cal Q}_B{\cal E})\|\le\|{\cal Q}_B{\cal E}\|_{2,2}^2\le
\|{\cal E}\|_{2,2}^2\le 1$  (where we have used that ${\cal Q}_B$ is a Hilbert-Schmidt projection and therefore $\|{\cal Q}_B\|_{2,2}\le1$).
Since $\hat{C}({\cal F})\ge 0$ this implies $0\le \hat{C}({\cal F})\le  \openone$.
{\bf{v)}} The first equality follows simply from $\hat{C}\ge 0$. Now, ${\mathrm{Tr}}\, \hat{C}({\cal E})={\mathrm{Tr}}\,\hat{A}({\cal E})-{\mathrm{Tr}}\,\hat{A}({\cal D}_B{\cal E})=
\sum_{i=1}^d (\|{\cal E}(P_i)\|_2^2-\|{\cal D}_B{\cal E}(P_i)\|_2^2)$. From this last expression, since for any state $\rho$ one has that $1/d\le \|\rho\|_2^2\le 1$, the desired inequality follows.
$\hfill\Box$\vskip 0.2truecm

\vskip 0.2truecm
{\bf{Remark 3}}
From the proof of {\bf{i)}} above and the remarks after Def. 2 one sees that in the space of normal unital maps a null coherence matrix is equivalent to incoherence i.e., the associated CGP measures are  {{faithful}}.
\vskip 0.2truecm

Any positive linear functional $p\colon M(d, {\mathbf{R}})\mapsto{\mathbf{R}}$ fulfills the properties required in {\bf{i)}}  above.
Important examples of the unitary invariant monotonic function $p$ in {\bf{i)}} are provided by the operator and trace norms over  $M(d, {\mathbf{R}}).$
We also note that applying the operational Protocol illustrated above to both operations ${\cal F}={\cal E}$ and ${\cal F}={\cal D}_B {\cal E}$
provides a direct way to determine experimentally the entries of the coherence matrix $\hat{C}({\cal E})$.


\section{CGP of quantum unitary maps}
In this section we will specialize to unitary operations ${\cal U}(X)=UXU^\dagger,\,(U\in U({\cal H})).$ In this case the coherence matrix simplifies
and a key element of the formalism becomes a bi-stochastic matrix $\hat{X}(U)\in M({\mathbf{R}}, d)$ that provides the $I_B$ simplex
representation of ${\cal U}$.
\vskip 0.2truecm
{\bf{Proposition 3}}

{\bf{i)}} $\hat{C}({\cal U})=\openone-\hat{B}({\cal U})$ where $\hat{B}({\cal U}):=\hat{X}^T(U)\hat{X}(U)$ with
\begin{equation}
\hat{X}(U)_{ij}:=\langle P_j, {\cal U}(P_j)\rangle=|\langle i|U|j\rangle|^2\quad(i,j=1,\ldots,d)
\label{B-Matrix}
\end{equation}
and $\hat{X}(U)$ a $d\times d$ {\em{bi-stochastic}} matrix.

{\bf{ii)}}
$\|\hat{C}({\cal U})\|=
1-\min_{\|v\|=1}\|\hat{X}(U)v\|^2.
$

{\bf{iii)}}
$\|\hat{C}({\cal U})\|_1=d-\|\hat{X}(U)\|_2^2.$
\vskip 0.2truecm
{\em{Proof.--}}
{\bf{i)}}
For any unitary $\hat{A}({\cal U})_{ij}=\langle {\cal U}(P_i), {\cal U}(P_j)\rangle=\langle P_i, P_j\rangle=\delta_{ij}\Rightarrow \hat{A}({\cal U})=\openone$.
Moreover, $\hat{A}({\cal D}_B{\cal U})_{ij}=\langle {\cal D}_B{\cal U}(P_i), {\cal D}_B{\cal U}(P_j)\rangle=\sum_{h, k=1}^d \langle P_k  {\cal U}(P_i) P_k, P_h  {\cal U}(P_j) P_h\rangle=
\sum_{k=1}^d  {\mathrm{Tr}}\left( P_k {\cal U}(P_i) P_k {\cal U}(P_j) \right)=\sum_{k=1}^d |\langle k|U|i\rangle|^2  |\langle k|U|j\rangle|^2=\sum_{k=1}^d \hat{X}^T(U)_{ik} \hat{X}(U)_{kj}=
(\hat{X}^T(U) \hat{X}(U))_{ij}$. Using Eq.~(\ref{C-Matrix}) completes the proof (bi-stochasticity of $\hat{X}(U)$ follows immediately from unitarity of $U$).
{\bf{ii)}} $\|\hat{C}({\cal U})\|=\max_{\|v\|=1}\langle v, \hat{C}({\cal U}) v\rangle= 1-\min_{\|v\|=1}\langle v, {\hat{B}}({\cal {U}}) v\rangle=1-\min_{\|v\|=1}\|\hat{X}(U)v\|_2^2$.
{\bf{iii)}} Using {\bf{v)}} of Prop. 2 it is immediate from {\bf{i)}} by taking the trace.
$\hfill\Box$\vskip 0.2truecm

It is well-know that any such bi-stochastic  matrix $X$ can be written as a convex combination of permutation matrices $X=\sum_{\sigma\in{\cal S}_d} q_\sigma P_\sigma$
where $P_\sigma=\sum_{l=1}^d |\sigma(l)\rangle\langle l|$ (notice that $\|P_\sigma\|^2_2={\mathrm{Tr}}(P_\sigma^T P_\sigma)=d$).
From convexity
$\|X\|_2^2=\|\sum_{\sigma\in{\cal S}_d} q_\sigma P_\sigma\|_2^2\le \sum_{\sigma\in{\cal S}_d} q_\sigma \|P_\sigma\|_2^2= d \sum_{\sigma\in{\cal S}_d} q_\sigma=d$
and the equality is fulfilled iff $X$ is a permutation matrix.
From this perspective we see that the {{CGP measure in {\bf{iii)}} is related to the ``distance" between $\hat{X}(U)$ and the  permutations i.e., the image under $\hat{X}$ of the incoherent unitaries}}.
Let's try to make this remark more precise.

\vskip 0.2truecm
{\bf{Proposition 4}}
\begin{equation}
{\bf{i)}}\quad\tilde{C}(U):=\min_{{\sigma\in{\cal S}_d}} \|\hat{X}(U)-P_\sigma\|^2_2\le \|\hat{C}({\cal U})\|_1\le d-1 \label{lower-bound}
\end{equation}
moreover $\tilde{C}(U)$ is itself a good CGP measure.

{\bf{ii)}} The two CGPs in Eq.~(\ref{lower-bound}) saturate the upper bound in (\ref{lower-bound}) over the same set of unitaries.
\vskip 0.2truecm
{\em{Proof.--}}
{\bf{i)}} The last upper bound is just a special case of {\bf{v)}} of Prop. 2. Let us  now consider
$\min_{{\sigma\in{\cal S}_d}} \|X-P_\sigma\|^2_2=\|X\|_2^2+d-2\max_{{\sigma\in{\cal S}_d}}\langle P_\sigma, X\rangle.$
If $Y=\sum_{\sigma\in{\cal S}_d} q_\sigma P_\sigma$ one has $\langle Y, X\rangle \le \max_{{\sigma\in{\cal S}_d}}\langle P_\sigma, X\rangle$.
Therefore for any bi-stochastic $Y$ one has
$\|X\|_2^2+d-2\langle Y, X\rangle\ge \min_{{\sigma\in{\cal S}_d}} \|X-P_\sigma\|^2_2,$ in particular setting $Y=X$ one finds $d-\|X\|_2^2\ge \min_{{\sigma\in{\cal S}_d}} \|X-P_\sigma\|^2_2$
from which Eq.~(\ref{lower-bound}) follows.
Also, if $W$ is incoherent then $\hat{X}(W)=P_{\sigma_W}$ and $\hat{X}(WU)=P_{\sigma_W} \hat{X}(U)$ and $\hat{X}(UW)= \hat{X}(U)P_{\sigma_W}$. 
Whereby $\|\hat{X}(WU)-P_\sigma\|_2=\|P_{\sigma_W} \hat{X}(U)-P_\sigma\|_2=\|\hat{X}(U)-P^{-1}_{\sigma_W}P_\sigma\|_2=\|\hat{X}(U)-P_{\sigma_W^{-1}\sigma}\|_2$
and therefore $\tilde{C}(WU)=\tilde{C}(U).$ Similarly $\tilde{C}(UW)=\tilde{C}(U)$ for any $U$ and incoherent $W$.
{\bf{ii)}} From {\bf{i)}} it follows that $\max_U \tilde{C}(U)\le \max_U \|\hat{C}({\cal U})\|$ so now we show that the two maxima are identical and achieved over the same set of unitaries.
In fact if $\hat{X}(U_*)_{ij}=1/d,\,\forall i,j$
one has that $\hat{X}(U_*)=|\phi^+\rangle\langle \phi^+|$ where $|\phi^+\rangle=d^{-1/2}\sum_{j=1}^d |j\rangle.$ Since now $\|\hat{X}(U_*)\|_2^2=\langle\phi^+|\phi^+\rangle=1$  it follows that
$\|\hat{X}(U_*)-P_\sigma\|_2^2=d+1-2{\mathrm{Tr}}(P_\sigma |\phi^+\rangle\langle \phi^+|)=d+1-2 \langle \phi^+|P_\sigma|\phi^+\rangle=d-1,\,\forall \sigma\in{\cal S}_d.$
Whereby $\tilde{C}(U_*)=d-1=\max_U \|\hat{C}({\cal U})\|_1$. Last equality holds because $\max_U  \|\hat{C}({\cal U})\|_1=d-\min_U\|\hat{X}(U)\|_2^2;$ but from bistochasticity of $\hat{X}(U)$
it follows that $\|\hat{X}(U)\|_2^2\ge 1,$  and $\|\hat{X}(U_*)\|_2^2=1$ which completes the proof.
$\hfill\Box$\vskip 0.2truecm
%

{\bf{Remark 4}} From {\bf{ii)}} of Prop. 3 one sees that $\|\hat{C}({\cal U})\|_1$ is a {\em{concave}} function of $\hat{X}(U)$ therefore its maximum is achieved
for the most ``mixed" element in the polytope of bi-stochastic matrices. This element is the uniform mixture of the extremal elements i.e., $(N!)^{-1}\sum_{\sigma\in{\cal S}_d} P_{\sigma}.$
In turn this element is clearly the projector over the trivial-irrep component of the ${\cal S}_d $ representation defined by the $ P_{\sigma}$'s.
It is easy to check that the only ${\cal S}_d$-invariant  vector is $|\phi^+\rangle.$ Whence the maximum of $\|\hat{C}({\cal U})\|_1$ is achieved  by those $U_*$'s such that $\hat{X}(U_*)=|\phi^+\rangle\langle \phi^+|$ as used in the above.

For example for $d=2$ if $\langle 0|U|0\rangle=  \langle 1|U|1\rangle^*=a,\,\langle 1|U|0\rangle=-\langle 0|U|1\rangle^*=b, \,(a,b\in{\mathbf{C}}, |a|^2+|b|^2=1)$ one finds that $\hat{X}(U)=|a|^2 \openone + |b|^2 \sigma_x$
and, by using the definition  in Eq.~(\ref{lower-bound}), that $$\tilde{C}(U)=4 \min \{|a|^4,\,|b|^4\}\le 4 |a|^2|b|^2=\|\hat{C}({\cal U})\|_1 .$$
Notice that these functions achieve the same maximum value $1$ when $|a|=|b|=1/\sqrt{2}.$

\subsection{CGP based on separation from incoherent maps}
In this  section we will generalize the geometrical approach to CGP hinted by Prop. 4. The idea is that a suitable notion of
geometrical separation e.g., a distance, of a map from the set of incoherent operations should provide a CGP.

Let $f$ a be map that takes super-operators to $\mathbf{R}_0^+$ such that:

{\bf{0)}} $f(X)=0$ iff $X=0;$ {\bf{1)}} $f(X)= f(-X);$ {\bf{ 2)}} $f({\cal W} X)=f(X{\cal W})=f(X)$ for all unitary ${\cal W}.$

For example $f$ could be any unitary invariant norm.
One can define a CGP for unitary CP maps ${\cal U}$ associated with $f$ in the following way (we assume finite dimensions)

{\bf{Definition 6}}
We define the following two non-negative valued functions over unitaries
\begin{eqnarray}
\tilde{C}_f( {\cal U}) &:=&\min_{{\cal W}\in CP_B} f({\cal W}{\cal D}_B-{\cal U}{\cal D}_B) \label{tildeC_f}\\
C_f({\cal U})&:=& f({\cal Q}_B{\cal U}{\cal D}_B) \label{C_f}
\end{eqnarray}
here the minimization is over incoherent unitaries i.e., such that ${\cal W}{\cal D}_B={\cal D}_B{\cal W}{\cal D}_B.$
\vskip 0.2truecm
{\bf{Proposition 5}} 

{\bf{i)}} $\tilde{C}_f({\cal U})\le C_f({\cal U})$.

{\bf{ii)}} $C_f$ and $\tilde{C}_f$  are good measures of CGP.

\vskip 0.2truecm
{\em{Proof.--}}
{\bf{i)}} Notice that ${\cal D}_B{\cal U}{\cal D}_B\in CP_B$ and therefore
$\tilde{C}_f({\cal U})\le f({\cal D}_B {\cal U}{\cal D}_B-{\cal U}{\cal D}_B)=C_f({\cal U})$.
{\bf{ii)}} Now from {\bf{0)}} above one has that $ {\cal Q}_B{\cal U}{\cal D}_B=0\Leftrightarrow C_f({\cal U})=0\Rightarrow  {\tilde{C}}_f({\cal U})=0.$ On the other hand if ${\tilde{C}}_f({\cal U})=0$ there exists  ${\cal W}_0\in CP_B$ such that
${\cal W}_0{\cal D}_B={\cal U}{\cal D}_B.$ Multiplying by ${\cal D}_B$ one gets ${\cal D}_B{\cal W}_0{\cal D}_B={\cal D}_B{\cal U}{\cal D}_B$ but one also has ${\cal D}_B{\cal W}_0{\cal D}_B={\cal W}_0{\cal D}_B= {\cal U}{\cal D}_B$ thereby ${\cal U} {\cal D}_B= {\cal D}_B{\cal U}{\cal D}_B$ from which $C_f({\cal U})=0.$
 We therefore see that $\tilde{C}_f({\cal U})=0\Leftrightarrow {\cal Q}_B{\cal U}{\cal D}_B=0.$ This shows that both functions vanish iff ${\cal U}\in CP_B$.
One can easily check that  both functions $C_f$ and $\tilde{C}_f$ fulfill the invariance property: ${\cal W}\in CP_B$ implies $ F({\cal W} {\cal U})= F({\cal U}{\cal W})=F({\cal U}),\,F=C_f, \tilde{C}_f.$  For example
$\tilde{C}_f({\cal W}{\cal U})= \min_{{\cal W}^\prime\in CP_B} f( {\cal W}^\prime{\cal D}_B-{\cal W}{\cal U}{\cal D}_B)=  \min_{{\cal W}^\prime\in CP_B} f( {\cal W}^{-1}{\cal W}^\prime{\cal D}_B-{\cal U}{\cal D}_B)=\tilde{C}_f({\cal U})$ where we have used {\bf{2)}} above and  that ${\cal W}^{-1}{\cal W}^\prime$ still runs across the whole $CP_B$ when $\cal W$ does.  $\hfill\Box$\vskip 0.2truecm

Let  $\||\bullet\||$ be any unitary invariant operator norm. For any ${\cal E}\in CP$ we define
\begin{equation}
f({\cal E})=\int d\mu(\psi)\, \||{\cal E}(|\psi\rangle\langle\psi|)\|| \label{norm}
\end{equation}
where the integration is over the Haar measure $d\mu(\psi)$. Notice that for any $\psi$ the map $p_\psi: {\cal E}\mapsto \||{\cal E}(|\psi\rangle\langle\psi|)\||$ is a (super-operator) semi-norm
and that the family $\{p_\psi\}_\psi $ is a {\em{sufficient}} family of semi-norms i.e., $X=0\Leftrightarrow p_\psi(X)=0 \, (\forall\psi)$.
 Eq.~(\ref{norm}) defines a unitary invariant norm of super-operators. Therefore $f$  fulfills properties {\bf{0)--2)}} above.
Moreover, from concavity one has $f({\cal E})\le \sqrt{g({\cal E})}$ where
\begin{equation}
g({\cal E}):=\int d\mu(\psi)\, \||{\cal E}(|\psi\rangle\langle\psi|)\||^2.
\label{g}
\end{equation}
One can check that the function ${\cal E}\mapsto {g({\cal E})}$ also fulfills the properties {\bf{0)--2)}} above.
In summary, by using Prop.5, we have proven the following:
\vskip 0.2truecm
{\bf{Proposition 6}} 

{\bf{i)}} The map $C_g\colon {\cal U}\mapsto g({\cal Q}_B{\cal U}{\cal D}_B)$ where $g$ is given by Eq.~(\ref{g}) is a good measure of CGP.

{\bf{ii)}} $\tilde{C}_f( {\cal U})\le {C}_f( {\cal U})\le\sqrt{C_g({\cal U})}$.
Moreover $\tilde{C}_f( {\cal U})=0\Leftrightarrow {C}_f( {\cal U})=0\Leftrightarrow C_g({\cal U})=0.$
\vskip 0.2truecm
The approach we  pursued in \cite{CGP_0}  amounts  to pick $\||X\||=\|X\|_2:=\sqrt{\mathrm{Tr}(X^\dagger X)}$ in Eq.~(\ref{norm}) and using the associated function $C_g$ as CGP. This can also be directly seen from Eq.~(\ref{CGP-Haar}).

\section{The Simplex Correlation Matrix}
In this section we define a second basic building block to construct families of CGP measures for unital quantum operations.
The underlying idea is simple: one prepares an input distribution $\mu$ of incoherent states in $\rho\in I_B,$ transforms them by the CP maps $\cal E$
and defines an associated CGP by the average coherence, as measured by $\|{\cal Q}_B{\cal E}(\rho)\|_2^2,$ generated by this stochastic process.
This provides a generalization of the approach pursued in \cite{CGP_0} where $\mu$ was always assumed to be the uniform measure over
$I_B$.
\vskip 0.2truecm
{\bf{Definition 7}}

{\bf{a)}} Given a probability distribution $\mu$  over $I_B$ with expectation ${\mathbf{E}}_\mu[\varphi]:=\int_{I_B} d\mu(x)  \varphi(x)$
($\varphi\colon I_B\mapsto {\mathbf{R}}$)  we  define the Simplex Correlation Matrix (SCM)
\begin{equation}
S_{ij}(\mu):={\mathbf{E}}_\mu[p_ip_j]\label{SCM}
\end{equation}
{\bf{b)}} We define the function
\begin{equation}
C_{B,\mu}({\cal E}):=\sum_{i,j=1}^d \hat{C}_{ij}({\cal E})  {\mathbf{E}}_\mu[p_i p_j] =\langle \hat{C}({\cal E}),\hat{S}(\mu)\rangle
\label{CGP-mu}
\end{equation}
\vskip 0.2truecm
{\bf{Example 0}} Eq.~(\ref{CGP-mu}) is of course a generalization of Eq.~(\ref{C-S}).
If $\mu_{Haar}$ is the uniform measure on $I_B$  one has that $S_{ij}(\mu_{Haar})=[d(d+1)]^{-1}(1+\delta_{ij})$ or in operator notation \cite{CGP_0}
\begin{equation}
\hat{S}({\mu_{Haar}})= \frac{1} {d(d+1)} (\openone + d |\phi^+\rangle\langle \phi^+|).
\label{SCM:Haar}
\end{equation}
In this case $C_{B,\mu_{Haar}}$ will be simply denoted by $C_{B}$ and coincides with the measure (\ref{CGP-Haar}) that we analyzed in depth in \cite{CGP_0}.

{\bf{Example 1}} If $\mu(p)=d^{-1}\sum_{i=1}^d \delta(p-e_i)$ where $(e_i)_l=\delta_{il}$ then $S(\mu)=d^{-1}\openone$.
Therefore the probability distribution is concentrated on the extremal points of $I_B$ i.e., the representatives of the basis projections $P_i$'s.

\vskip 0.2truecm
{\bf{Proposition 7}}

{\bf{i)}} $\hat{S}(\mu)=\hat{S}(\mu)^T\ge 0$  and $\sum_{i,j=1}^d \hat{S}_{ij}(\mu)= d\langle\phi^+|\hat{S}(\mu)|\phi^+\rangle=1$.

{\bf{ii)}} The function $ C_{B,\mu}({\cal E})$ defined by Eq.~(\ref{CGP-mu}) is a good CGP measure.

{\bf{iii)}} If $s_M(\mu)$ ($s_m(\mu)$) is the maximum (minimum) eigenvalue of the SCM $\hat{S}(\mu)$ then
\begin{equation}
s_m(\mu) {\mathrm{Tr}}\, \hat{C}({\cal E}) \le C_{B,\mu}({\cal E})\le s_M(\mu) {\mathrm{Tr}}\, \hat{C}({\cal E}).
 \label{C_{B,mu}}
\end{equation}


{\em{Proof.--}}
{\bf{i)}} $\hat{S}(\mu)_{ij}={\mathbf{E}}_\mu[p_i p_j]={\mathbf{E}}_\mu[p_j p_i]=\hat{S}(\mu)_{ji},\,(\forall i,j)$;
$\sum_{i,j=1}^d v_i v_j (\hat{S}_\mu)_{ij}= {\mathbf{E}}_\mu[(\sum_{i=1}^d v_i p_i)^2]\ge 0.$
Moreover $\sum_{i,j=1}^d \hat{S}_{ij}(\mu)={\mathbf{E}}_\mu[(\sum_{i=1}^d p_i)^2]={\mathbf{E}}_\mu[1]=1$.
{\bf{ii)}} Since $\hat{S}(\mu)\ge 0$ the function $p\colon\hat{C}\mapsto {\mathrm{Tr}} (\hat{C} \hat{S}(\mu))$ is a positive
linear functional. The claim then follows from {\bf{i)}} of Prop. 2.
{\bf{iii)}} Obvious from Eq.~(\ref{CGP-mu}) and the spectral representation of $\hat{S}(\mu)$. $\hfill\Box$
\vskip 0.2truecm
For the Haar's measure from Eq.~(\ref{SCM:Haar}) one has that $C_B({\cal E})=[d(d+1)]^{-1} {\mathrm{Tr}}\, \hat{C}({\cal E})$
(remember $\langle\phi^+|\hat{C}({\cal E})|\phi^+\rangle=0$).
In particular in the  unitary case $ {\mathrm{Tr}}\, \hat{C}({\cal U})=d-\|\hat{X}(U)\|_2^2$ (see above).
If SCM is full rank i.e., $s_m(\mu)>0,$ from (\ref{C_{B,mu}}) one has that the CGP with respect to the distribution $\mu$ vanishes iff the Haar's measure CGP does. Moreover
$$s_m(\mu) d(d+1)  C_B({\cal E}) \le C_{B,\mu}({\cal E})\le s_M(\mu)  d(d+1)  C_B({\cal E}).$$
For the unitary case, using Eqs.~(\ref{lower-bound}) and (\ref{C_{B,mu}}), we have that, for any full rank $\mu$,
\begin{equation}
s_m(\mu) \min_{{\sigma\in{\cal S}_d}} \|\hat{X}(U)-P_\sigma\|^2_2\le  C_{B,\mu}({\cal U}).
\label{lower-bound-mu}
\end{equation}

\subsection{Permutational Invariance}
The definition of CGP requires invariance under post-processing by incoherent unitaries $W$. We now analyze under  which conditions invariance holds for pre-processing
as well.
If ${\cal W}$ is an incoherent unitary CP then $\hat{X}({\cal W})=\sum_{i=1}^d  |\sigma_W(i)\rangle\langle i|=:P_{\sigma_W},\,\sigma_W\in{\cal S}_d$ \cite{CGP_0}.
\vskip 0.2truecm
{\bf{Proposition 8}}

{\bf{i)}} $\hat{C}({\cal U}{\cal W})= P_{\sigma_W}^T \hat{C}({\cal U})P_{\sigma_W}$

{\bf{ii)}} $C_{B,\mu}({\cal U}{\cal W})=C_{B,\mu\circ  P_{\sigma_W}^{-1}}({\cal U})$.


{\em{Proof.--}}
{\bf{i)}} It follows from {\bf{i)}} of Prop. 3 and the fact $\hat{X}({\cal U}{\cal W})=\hat{X}({\cal U}) P_{\sigma_W}$

{\bf{ii)}} From the above
\begin{eqnarray}
C_{B,\mu}({\cal U}{\cal W})=\langle P_{\sigma_W}^T \hat{C}({\cal U})P_{\sigma_W}, \hat{S}(\mu)\rangle= \langle \hat{C}({\cal U}), \hat{S}^W(\mu) \rangle\nonumber
 \end{eqnarray}
 where $ \hat{S}^W(\mu) :=P_{\sigma_W} \hat{S}(\mu) P_{\sigma_W}^T$.

 Now, $\hat{S}^W(\mu) =\sum_{l,m} (P_{\sigma_W})_{il} {\mathbf{E}}_\mu[p_l p_m] (P_{\sigma_W})^T_{mj}= {\mathbf{E}}_\mu[ (\sum_l (P_{\sigma_W})_{il} p_l) (\sum_m (P_{\sigma_W})_{jm} p_m)]=
 {\mathbf{E}}_\mu[ q_i q_j],$ where $q=P_{\sigma_W}(p)\in I_B.$ By denoting with $f_{ij} (p)$ the argument of the expectation $ {\mathbf{E}}_\mu$ one has that
$S^W(\mu)_{ij}={\mathbf{E}}_\mu[f_{ij}(P_{\sigma_W}(p))]=\int_{I_B}d\mu(p) f_{ij}(P_{\sigma_W}(p))=\int_{I_B}d\mu(P_{\sigma_W}^{-1}(q)) f_{ij}(q).$
If $d\mu(p)= g(p) d^np$ we have $d\mu(P_{\sigma_W}^{-1}(q))=g(P_{\sigma_W}^{-1}(q)) d^nq $ where we used $|{\mathrm{det}}P_{\sigma_W}|=1.$
This shows that $\hat{S}^W(\mu)=\hat{S}(\mu\circ  P_{\sigma_W}^{-1}).$ $\hfill\Box$
\vskip 0.2truecm
{\bf{Remark 5}} In general $\mu\neq \mu\circ  P_{\sigma_W}^{-1}$ and therefore $C_{B,\mu}({\cal U}{\cal W})\neq C_{B,\mu}({\cal U})$ i.e., in general invariance by pre-processing with an incoherent unitary does not hold. A sufficient condition for invariance is that $S^W(\mu)=\hat{S}(\mu)$ for all incoherent $W$'s, namely $[\hat{S}(\mu), P_\sigma]=0,\,\forall\sigma\in{\cal S}_d.$
For example this condition is (obviously) fulfilled by the Haar's measure SCM Eq.~(\ref{SCM:Haar}). If $S$ commutes with all the $P_\sigma$'s then $S=\alpha {\openone}+\beta |\phi^+\rangle\langle\phi^+|$, whereas positivity and $\sum_{ij}S_{ij}=1$ imply $\alpha\ge 0$ and $\beta=1/d-\alpha$.
From this, using $\langle\phi^+|\hat{C}({\cal E})|\phi^+\rangle=0$ for all unital $\cal E$'s, one finds the general form of the CGP for {\em{permutationally invariant}} $\mu$
\begin{equation}
C_{B,\mu}({\cal E})= \alpha(\mu)  {\mathrm{Tr}}\,\hat{C}({\cal E}),\quad 0\le\alpha(\mu)\le 1/d. \label{CP-perm-inv}
\end{equation}
Here $\alpha(\mu)={\mathbf{E}}_\mu[p_i^2]-{\mathbf{E}}_\mu[p_l p_k],\,(k\neq l)$.
The Haar's case corresponds to $\alpha(\mu_{Haar})^{-1}=d(d+1).$
From Eq.~(\ref{CP-perm-inv}) one finds that the maximum of CGP for permutation invariant measures is obtained by $\alpha=1/d$ ($\Rightarrow\beta=0)$.
In words:
for any $U$ maximal CGP  (over permutational invariant $\mu$'s) is achieved for the $\mu$ that is uniformly  concentrated over the extremal points of $I_B.$
This result is not surprising as the CGP defined in Eq.~(\ref{CGP-mu}) is the expectation of a convex function that achieves its maximum over the vertices of $I_B.$
The minimum of (\ref{CP-perm-inv}) is of course achieved by $\alpha=0$ which corresponds to a measure concentrated on the center of $I_B$ i.e., the maximally mixed state
${\openone}/d$.
\vskip 0.2truecm
\subsection{Qubit Case}
Let us illustrate these findings in the qubit case $d=2$
(notice that  $\alpha$ in this example corresponds to $\frac{1}{2}(\alpha+\frac{1}{2})$ of the lines above).
For $d=2$ the simplex is a segment $I_B=[0,1]\ni p$ and the invariance condition is fulfilled by any $d\mu(p)= g(p) dp $ such that $g(p)=g(1-p).$ One finds
$\hat{S}(\mu)=\alpha  {\openone} +\beta \sigma_x$ where $\alpha={\mathbf{E}}_\mu[p^2]={\mathbf{E}}_\mu[(1-p)^2]$ and $\beta={\mathbf{E}}_\mu[p(1-p)]\le \alpha$.
Since $\alpha+\beta=1/2$ (from $\sum_{i,j=1}^d S_{ij}=1$) we have that $1/4\le \alpha,\beta\le 1/2.$
Moreover, if $U=a|0\rangle \langle 0| +a^* |1\rangle  \langle 1|-b^*|0\rangle\langle 1| +b |1\rangle\langle 0|$ ($|a|^2+|b|^2=1$) then
$\hat{X}(U)=|a|^2 {\openone} +|b|^2\sigma_x\Rightarrow \hat{C}(U)={\openone}-\hat{X}(U)^2=2|a|^2|b|^2 (\openone-\sigma_x)=4|a|^2|b|^2 |-\rangle\langle-|.$
Thereby,
$$C_{B,\alpha}(U)={\mathrm{Tr}} \left(\hat{C}(U)\hat{S}(\alpha)\right)=(2\alpha-\frac{1}{2})(2|a||b|)^2.$$
(Haar's case is $\alpha=1/3$)
For $\alpha=1/4$ one has that the CGP is identically vanishing for {\em{all}} $U$'s, which corresponds to a $\mu$ concentrated over the centre of $I_B$
i.e., the maximally mixed state $\rho={\openone}/2$.  For  $\alpha=1/2$ one has maximal CGP. The latter case corresponds to $\mu$ being equally concentrated
on $p=0$ and $p=1$ i.e.,  $\mu(p)=1/2(\delta(p)+\delta(1-p))$

\section{CGP and Tensor  Product Structures}
In this section we will  discuss the relation between CGP measures and Tensor Product Structures (TPS).  The goal is to define CGP measures that are
compatible with this additional structure. This is the relevant situation for multi-partite quantum systems.
In particular we would like to introduce {\em{additive}} measures i.e., such that the CGP of a tensor product of operations
is the sum of the individual CGPs.

Let us assume
${\cal H}\cong{\cal H}_A\otimes{\cal H}_B$ where ${\cal H}_X={\mathrm{span}}\{ |i_X\rangle\}_{i_X=1}^{d_X},\,(X=A,B)$
and consider the product basis $A\otimes B := \{|i_A\rangle\otimes |i_B\rangle\}_{i_A,i_B}.$
First observe that ${\cal D}_{A\otimes B}={\cal D}_{A}\otimes {\cal D}_{B}$ from which follows that the
$\hat{B}_{i_Ai_B, j_A j_B} ({\cal E}):=\langle {\cal D}_{A\otimes B}{\cal E}(|i_A i_B\rangle\langle i_A i_B|),  {\cal D}_{A\otimes B}{\cal E}(|j_A j_B\rangle\langle j_A j_B|)\rangle$
tensor factorizes for ${\cal E}={\cal E}_A\otimes {\cal E}_B.$ In short
$\hat{B}({\cal E}_A\otimes {\cal E}_B) =\hat{B}({\cal E}_A)\otimes\hat{B}({\cal E}_B).
$
Moreover, for unitaries
\begin{equation}
\hat{X}(U_A\otimes U_B)=\hat{X}(U_A)\otimes \hat{X}(U_B).
\label{tensor}
\end{equation}
Since the coherence matrix of a unitary, at variance with the general unital case,  is a function of $\hat{X}(U)$ only  (See {\bf{i)}} of Prop. 3) the factorization property
(\ref{tensor}) is the key one which leads to additivity.
\vskip 0.2truecm
{\bf{Proposition 9}}

Consider the functions over unitaries $U\in U({\cal H}_A\otimes {\cal H}_B)$
\begin{eqnarray}
{\bf{i)}}\quad \varphi_p(U) &:=& - \log \frac{\|\hat{X}(U)\|_2^2}{d}\in[0,\log d] \label{varphi_p}
 \\
{\bf{ii)}}\quad {\varphi}_g(U)&:=&-\frac{1}{d} \log |\det\,\hat{X}(U)|\in[0, \infty] \label{varphi_g}
\\
{\bf{iii)}}\quad {\varphi}_{\tilde{g}}(U)&:=& -\frac{1}{d}\sum_{i,j=1}^d \hat{X}_{ij}(U)\log \hat{X}_{ij}(U)\in[0,\log d], \label{tildevarphi_p}
\end{eqnarray}
these are {\em{additive}}  CGP measures i.e., $\varphi_x(U_A\otimes U_B)=\varphi_x(U_A)+\varphi_x(U_B),\,(x=p,g, \tilde{g}).$
Moreover they all achieve their maximum for $\hat{X}(U)=|\phi^+\rangle\langle\phi^+|$.
\vskip 0.2truecm
{\em{Proof.--}}
For incoherent $W$'s the corresponding $\hat{X}(W)\in{\cal S}_d\Rightarrow \|\hat{X}(W)\|_2^2=d, \, |\det\,\hat{X}(W)|=1,\, \hat{X}(W)_{ij}\in\{0, 1\} \forall i, j$
for which it is immediate to see $\varphi_x(W)=0\,(x=p,g, \tilde{g}).$
Moreover, since $\hat{X}(WU)=P_{\sigma_W} \hat{X}(U)$ one can explicitly check that invariance for post-processing by incoherent unitaries $W$ holds in all three cases.
Additivity  of the measures follows from Eq.~(\ref{tensor}),  $d=d_A d_B$, $\|X\otimes Y\|_2= \|X \|_2 \|Y\|_2$, and  $\det(X\otimes Y)=\det(X)^{d_B} \det(Y)^{d_A}$.
Finally  $\hat{X}(U)=|\phi^+\rangle\langle\phi^+|\Rightarrow \|\hat{X}(U)\|_2^2=1,\,|\det\,\hat{X}(U)|=0,\,\hat{X}(U)_{ij}=1/d$
from which explicit computation shows that (\ref{varphi_p}), (\ref{varphi_g}) and (\ref{tildevarphi_p}) achieve their maxima.
$\hfill\Box$
\vskip 0.2truecm
{\bf{Remark 6}} By defining the basic probability vectors ${\mathbf{p}}_i^{U}:= \hat{X}(U) e_i,\,[(e_i)_l=\delta_{il},\, ({\mathbf{p}}_i^{U})_l= \hat{X}(U)_{li}]$
one can write
\begin{equation}
{\varphi}_p(U)=-\log(\frac{1}{d} \sum_{i=1}^d \| {\mathbf{p}}_i^{U}\|^2) \le -\frac{1}{d}\sum_{i}^d \log \| {\mathbf{p}}_i^{U}\|^2=:{\varphi}_{2}(U)
\end{equation}
where we have used convexity of $-\log$. The upper bound can we rewritten as ${\varphi}_{2}(U)=\frac{1}{d}\sum_{i}^d S_2({\mathbf{p}}_i^{U})$
in which we introduced the $2$-Renyi entropy $S_2({\mathbf{p}})=-\log \|{\mathbf{p}}\|^2$ for a probability vector ${\mathbf{p}}$.
These considerations naturally lead to the following generalization using $\alpha$-Renyi entropies.
\vskip 0.2truecm
{\bf{Proposition 10}}
Let us define
${\varphi}_\alpha(U):=\frac{1}{d}\sum_{i=1}^d S_\alpha({\mathbf{p}}_i^{U})$ where $S_\alpha({\mathbf{p}}):=\frac{1}{1- \alpha}\log  \sum_{l=1}^d p_l^\alpha.$
 The functions ${\varphi}_\alpha$'s are additive CGP measures for unitaries $\forall \alpha\in [0, 2]$.
 \vskip 0.2truecm
 {\em{Proof.--}} Reasoning as in Prop. 9 above. $\hfill\Box$

 \vskip 0.2truecm
{\bf{Remark 7}} For $U$ such that $\hat{X}(U)=|\phi^+\rangle\langle\phi^+|$ one has that $({\mathbf{p}}_i^{U})_l=1/d\, (\forall i, l)$ and this implies that
 ${\varphi}_\alpha(U)=\log d \,(\forall\alpha)$ i.e., the maximum is achieved.
Moreover, since $\lim_{\alpha\to 1}  S_\alpha({\mathbf{p}})=-\sum_{l=1}^d p_l\log p_l:=H({\mathbf{p}})$ one sees that
$${\varphi}_{\alpha\to 1}(U)={\varphi}_{\tilde{g}}(U)\ge {\varphi}_{2}(U)\ge {\varphi}_{p}(U)$$
 where we have used $\alpha\le \beta\Rightarrow S_\alpha\ge S_\beta,\,\alpha,\beta\in [0,2]$.
\section{Conclusions}
Coherence and its generation are key properties for a variety of quantum information processing and control strategies.
Given an orthonormal basis in  a $d$-dimensional Hilbert space
we have  defined  a non-linear mapping that associates to a unital operation $\cal E$  acting on $\cal H$ a  $d\times d$ real-valued matrix that we call the coherence matrix
of $\cal E$ with respect to $B$.

We have shown that one can exploit  this coherence  matrix to introduce  vast families of  measures of the coherence generating power (CGP) for a unital quantum operation.
Interestingly, these measures have a natural geometrical interpretation as separation of $\cal E$ from the set of incoherent unital operations.

By means of  the coherence matrix we have reformulated and generalized the probabilistic approach to CGP  discussed in  \cite{CGP_0}.
In order to achieve this goal we have introduced a  second  $d\times d$ real-valued matrix, the simplex correlation matrix, that encodes for the relevant statistical correlations
in the input ensemble of incoherent states.  Contracting these two matrices one obtains  CGP measures describing  the process of preparing
a given input  ensemble and processing it with the chosen unital operation.

In the important unitary case we have discussed how these tools   can be made compatible with an underlying tensor product structure
by defining families of CGP measures that are additive.

As in \cite{CGP_0} the main tools utilized in this paper rely on the Hilbert-Schmidt scalar product in the system operator space. On the one hand this
allows one to define CGP measures that are easy to compute in any finite dimension. On the other hand one is limited to the unital case.
Overcoming this limitation, as well as to generalize to infinite dimensions, are still open important challenges for future research.

\acknowledgements
This work was partially supported by the ARO MURI grant W911NF-11-1-0268 and W911NF-15-1-0582.



\begin{thebibliography}{10}
\bibitem{example-georgios} The condition ${\cal Q}_B {\cal E}{\cal D}_B=0$ does not, in general, imply
that $[{\cal D}_B,{\cal E}]=0$. Consider for example the qubit  CP-map $X\mapsto ({\mathrm{Tr}}\,X)\, \openone +\epsilon (X_{01}+X_{10})\sigma^z,\,\epsilon \in [-\frac{1}{2}, \frac{1}{2}]-\{0\}$.
\end{thebibliography}
\end{document}